\documentclass[10pt,letterpaper,twocolumn]{article} 

\usepackage{ol2}
\usepackage[draft]{hyperref}
\usepackage{amsmath}

\begin{document}

\twocolumn[ 

\title{Stochastic non-Hermitian skin effect}


\author{Stefano Longhi}
\address{Dipartimento di Fisica, Politecnico di Milano and Istituto di Fotonica e Nanotecnologie del Consiglio Nazionale delle Ricerche, Piazza L. da Vinci 32, I-20133 Milano, Italy (stefano.longhi@polimi.it)}
\address{IFISC (UIB-CSIC), Instituto de Fisica Interdisciplinar y Sistemas Complejos, E-07122 Palma de Mallorca, Spain}

\begin{abstract}
A hallmark of photonic transport in non-Hermitian lattices with asymmetric hopping is the robust unidirectional flow of light, which is responsible for important phenomena such as  the non-Hermitian skin effect. Here we show that the same effect can be induced by stochastic fluctuations in lattices which maintain a symmetric hopping on average. We illustrate such a fluctuation-induced non-Hermitian transport by discussing stochastic funneling of light, in which light is pushed toward an interface by the stochastic-induced skin effect.
\end{abstract}

 ] 


{\it Introduction.} Discretized photonic transport and light walks in non-Hermitian lattices show a wealth of interesting effects which have attracted a major relevance in  the past few years. In particular, lattices with asymmetric hopping display rather generally a unidirectional flow of light, which is responsible for such phenomena as non-Hermitian Anderson delocalization transition \cite{r1,r2,r3} and non-Hermitian transparency \cite{r4}, with interesting applications to e.g. robust light transport \cite{r5}, reflectionless transmission \cite{r6}, nonreciprocal photon transmission and amplification \cite{r6b,r7},  topological lasers \cite{r8},  laser array synchronization \cite{r10}, non-Hermitian interferometry \cite{r11}, funneling of light \cite{r12}, and vortex micolaser control \cite{r13}. On a more fundamental viewpoint, non-Hermitian light walks provide a suitable platform to observe a wealth of recently discovered phenomena like the non-Hermitian skin effect (NHSE) \cite{r14,r15,r16,r17}, i.e. the accumulation of all eigenvectors in a finite-sized system toward the edges, the breakdown of the bulk-boundary correspondence based on Bloch-band topological invariants \cite{r14,r16,r18,r19,r20,r21}, non-Bloch and dynamic phase transitions \cite{r17,r22,r23,r23b}. While the impact of static spatial disorder in such systems has been investigated, with the prediction of important effects as non-Hermitian delocalization transition \cite{r1,r2,r3,r12,r24}, non-Hermitian transparency \cite{r4} and Anderson skin effect \cite{r25}, the role of temporal fluctuations of spatially homogeneous systems remains unexplored. Since fluctuations can induced transport in non equilibrium systems \cite{r25b}, it is of major relevance to understand the interplay between stochastic fluctuations, non-Hermitian transport and the skin effect. \\ 
In this Letter we disclose a rather counterintuitive photonic transport mechanism in a lattice with spatially-ordered but {\em fluctuating} hopping amplitudes, dubbed stochastic non-Hermitian skin effect, for which 
unidirectional light flow arises as a result of fluctuations solely. The effect is illustrated by considering  funneling of light at an interface induced by stochastic fluctuations.\\
\\
  \begin{figure*}[htbp]
\centerline{\includegraphics[width=17cm]{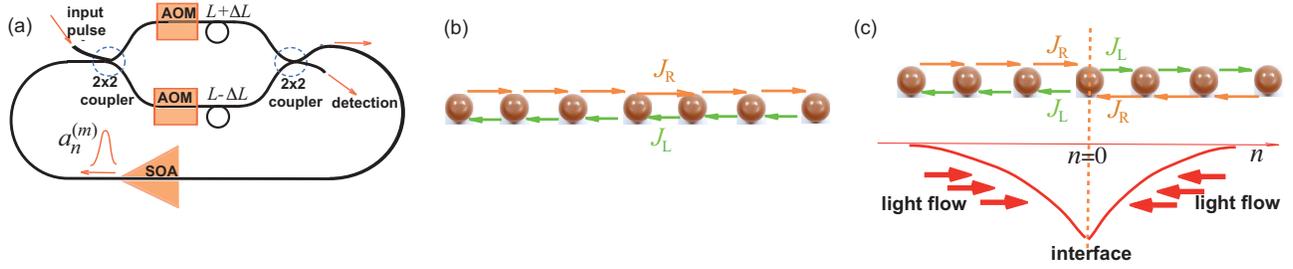}} \caption{ 
(Color online) (a) Schematic of  the fiber loop that realizes a synthetic mesh lattice \cite{r26}. An optical pulse propagating in the main fiber loop is periodically splitted and recombined in an interferometer with unbalanced arms of length $L \pm \Delta L$. Two acousto-optic modulators (AOMs) provide controllable loss in the two arms, while a semiconductor optical amplifier (SOA) in the main loop controls the overall gain per transit. The evolution of pulse amplitudes $a_n^{(m)}$ at successive transits $m$ in the loop is described by Eq.(1). The values of $J_R$ and $J_L$ can be tailored both in $n$ (fast time) and $m$ (slow time) by judicious control of loss rates in the two AOMs and gain in the SOA. (b) Schematic of a translational-invariant lattice with asymmetric hopping $J_{L,R}$. The continuous-time light walk on the lattice realizes the Hatano-Nelson model. Light flows unidirectionally for asymmetric hopping $J_L \neq J_R$. (c) Principle of light funneling based on the NHSE. An interface at $n=0$ is created by flipping left and right hopping amplitudes in the two half lines. For $J_R>J_L$, any initial excitation in the system flows toward the interface due to the skin effect.}
\end{figure*} 
{\it Discrete-time non-Hermitian light walk and the skin effect}. In order to illustrate the main idea of the present work, let us consider the discrete-time non-Hermitian light walk on a one-dimensional lattice with asymmetric hopping $J_{L,R}$ described by the following  map
\begin{equation}
-i a^{(m+1)}_{n}= J_R a^{(m)}_{n-1}+J_L a^{(m)}_{n+1}
\end{equation}
where $a_n^{(m)}$ is the light field amplitude at the site $n$ of the lattice and at the discrete time $m$. This model, earlier introduced in \cite{r26}, can be implemented in a fiber optics setup [Fig.1(a)],  where periodic circulation of optical pulses in fiber loops realizes a synthetic mesh lattice \cite{r27,r28,r29,r30,r31,r32}, and can be regarded as a discrete-time version of the Hatano-Nelson model [Fig.1(b)]. Like the Hatano-Nelson model \cite{r1,r2,r3,r4,r25},  for asymmetric hopping $J_L \neq J_R$ distinct quasi-energy spectra are found for a lattice with periodic (PBC) or open (OBC) boundary conditions. In the former case the quasi energy spectrum describes a closed loop complex plane and unidirectional light flow is observed in the bulk of the lattice, while in the latter case the quasi energy describes an arc and all eigenstates are squeezed toward the edge of the lattice (skin effect).  In fact, the quasi-energy spectrum $\mu$ is obtained by introducing the Ansatz $a_n^{(m)}=\exp(-i \mu m) A_n$ in Eq.(1), where $\mu$ and corresponding eigenvectors $A_n$ are obtained from the Hatano-Nelson eigenvalue problem
\begin{equation}
\exp(-i \mu -i \pi/2) A_n=J_R A_{n-1}+J_L A_{n+1}
\end{equation}
with appropriate boundary conditions. Under PBC, $A_n= \exp(ikn)$ where  $-\pi < k \leq \pi$ is the Bloch wave number, and the quasi-energy spectrum reads
\begin{equation}
\mu(k)=-\frac{\pi}{2}+i \log \left\{ J_L \exp(ik)+J_R \exp(-ik) \right\}
\end{equation}
  Note that the largest growth rate, i.e. largest imaginary part of $\mu$, is given by $\log (J_L+J_R)$ and is attained at $k=0, \pi$. The corresponding group (drift) velocity 
    \begin{equation}
  v_g=\left( \frac{d {\rm {Re}} (\mu) } {dk} \right)_{0, \pi}= \frac{J_R-J_L}{J_R+J_L}
  \end{equation}
   is non-vanishing and changes sign at the symmetric (phase transition) point $J_R=J_L$. According to the general results of Ref.\cite{r23}, this means that under PBC a robust unidirectional flow of light is expected at a drift velocity $v_g$, while the skin effect is observed under OBC. In a lattice comprising $N$ sites with OBC the quasi-energy spectrum is obtained from Eq.(3) after the replacement $k \rightarrow  
  k-ih$, where $h= (1/2) {\rm log} (J_R / J_L)$ is the complex shift of quasi-momentum and $k$ is quantized according to $k=k_l= l \pi/(N+1)$ ($l=1,2,...,N$), corresponding a deformation of the Brillouin zone $ \beta \equiv \exp(ik)$ from the unit circle $|\beta|=1$ under PBC to the circle $|\beta|=R$ of radius $R=\exp(h)$ under OBC \cite{r14,r15,r23}. The corresponding eigenvectors, given by
  \begin{equation}
  A_n= \sin (n k_l) \exp(hn)
  \end{equation}
  are squeezed toward either the left ($h<0$) or right ($h>0$) edge of the lattice. A typical behavior of quasi-energies for either PBC or OBC is shown in Fig.2.\\
The skin effect can be observed in the infinite lattice when an interface is introduced at site $n=0$ by flipping left and right hopping amplitudes in two half-lines, as shown in Fig.1(c). For $J_R>J_L$, all eigenstates condensate near $n=0$ and an arbitrary initial excitation of the system is pulled toward the interface (Fig.3), an effect dubbed light funneling \cite{r12}.\\
 
  \begin{figure}[htbp]
\centerline{\includegraphics[width=8.7cm]{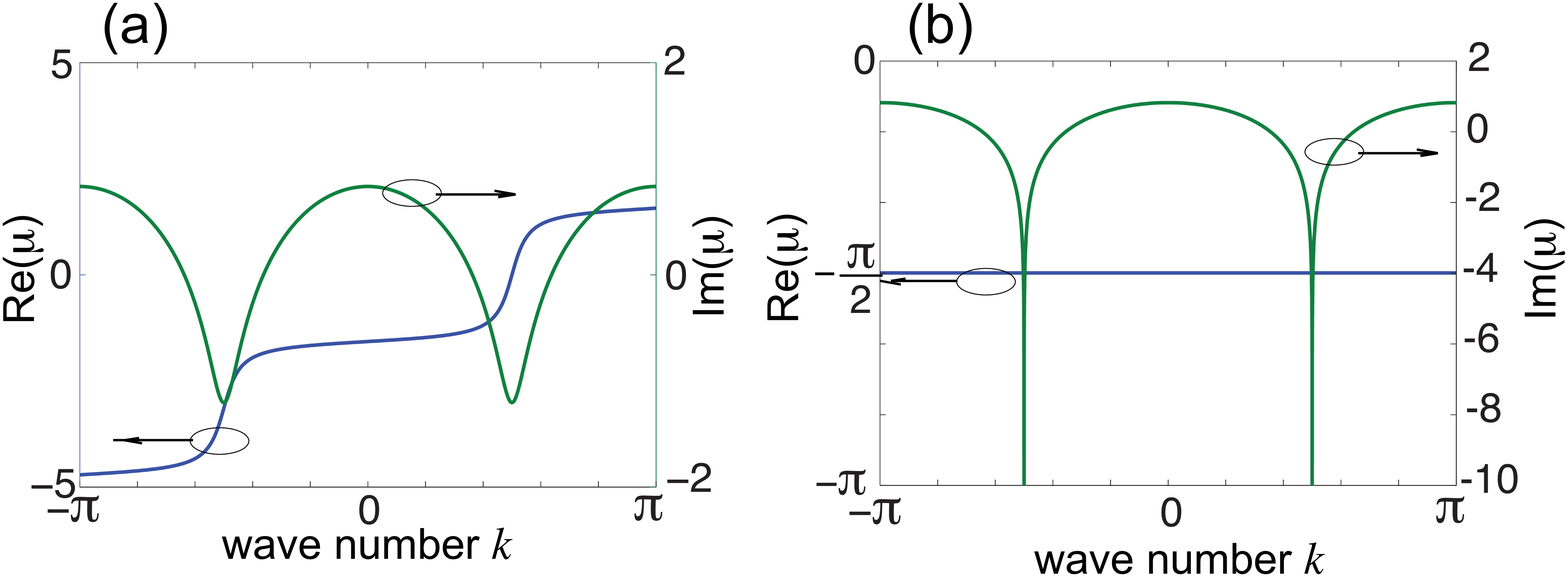}} \caption{ 
(Color online) (a) Quasi-energy spectrum (real and imaginary parts of the quasi energy $\mu$) of the discrete system (1) for $J_L=1$, $J_R=1.3$ and for (a) PBC, and (b) OBC.}
\end{figure} 

{\em Stochastic non-Hermitian skin effect and light funneling.} Let us now consider a lattice in which one of the two hopping amplitudes, for example $J_R$, changes stochastically at successive time steps $m$ in such a way that, on average, $J_L=J_R=J$, i.e.  let us assume in Eq.(1)
 \begin{equation}
 J_L=J \; , \; \; J_R= J(1+\xi^{(m)})
 \end{equation}
 where $J$ is the (symmetric) hopping amplitude in the absence of fluctuations and $\xi^{(m)}$ is a set of independent random variables with the same probability density distribution $f(\xi)$ of zero mean, i.e. $\langle \xi \rangle = \int \xi f(\xi) d \xi =0$. Clearly, in the absence of fluctuations there is not unidirectional light flow on the lattice (the drift velocity $v_g$ vanishes) and the skin effect is absent (since $h=0$). One might naively expect that, for a fluctuating hopping amplitude $J_R$ such that {\em on average} the left/right hopping amplitudes are equal, nothing will change. However, such a prediction is not correct and, rather counterintuitively, a fluctuation can produce a unidirectional light flow and the skin effect despite the hopping is symmetric on average. In fact, let us consider an infinitely extended lattice, so that any initial excitation of the system can be written as a superposition of Bloch states as
 \begin{equation}
 a_n^{(0)}= \int_{-\pi}^{\pi} dk F(k) \exp(ikn)
 \end{equation}
 with some spectral amplitude $F(k)$. From Eqs.(1), (6) and (7) it can be readily shown that the amplitudes $a_n^{(m)}$ at successive time steps $m$ read
 \begin{equation}
 a_n^{(m)}= \int_{-\pi}^{\pi} F(k)  \exp \left\{ ikn-i \mu_m(k) m \right\}
 \end{equation}
 where we have set
 \begin{figure}[htbp]
\centerline{\includegraphics[width=8.7cm]{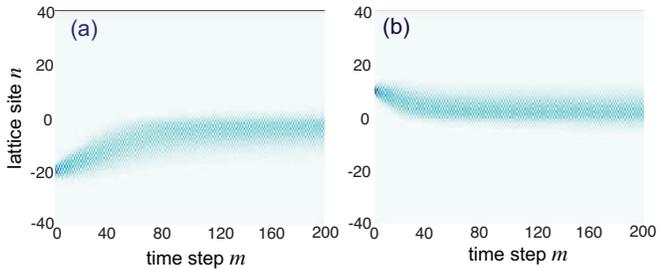}} \caption{ 
(Color online) Funneling of light at the interface of Fig.1(c) for $J_R=1.5$ and $J_L=1$. In (a) the lattice is excited in the single site $n=-20$, wheres in (b) the lattice is excited in the single site $n=10$. The numerically-computed behavior of the normalized light intensity $|a_n^{(m)}|^2 / \sum_n |a_n^{(m)}|^2$ versus time step $m$ is displayed on a pseudo color map.}
\end{figure}  \begin{equation}
 \mu_m(k)=- \frac{\pi}{2} +\frac{i}{m} \sum_{l=0}^{m-1} \log \left[ J \exp(ik)+ J(1+\xi^{(l)}) \exp(-ik) \right] .
 \end{equation}
In the long time limit $m \rightarrow \infty$, $\mu_m(k)$ converges to the stationary value (Lyapunov exponent)
\begin{equation}
 \mu (k)=- \frac{\pi}{2} +i \int_{-\infty}^{\infty} d \xi f( \xi) \log \left[ J \exp(ik)+ J(1+\xi) \exp(-ik) \right] 
\end{equation}
  which provides a generalization of the concept of quasi energy $\mu(k)$ [Eq.(3)] to the time-dependent (fluctuating) case. For example, assuming that $\xi^{(m)}$ is uniformly distributed in the range $(-\Delta, \Delta)$, one obtains
    \begin{figure}[htbp]
\centerline{\includegraphics[width=8.7cm]{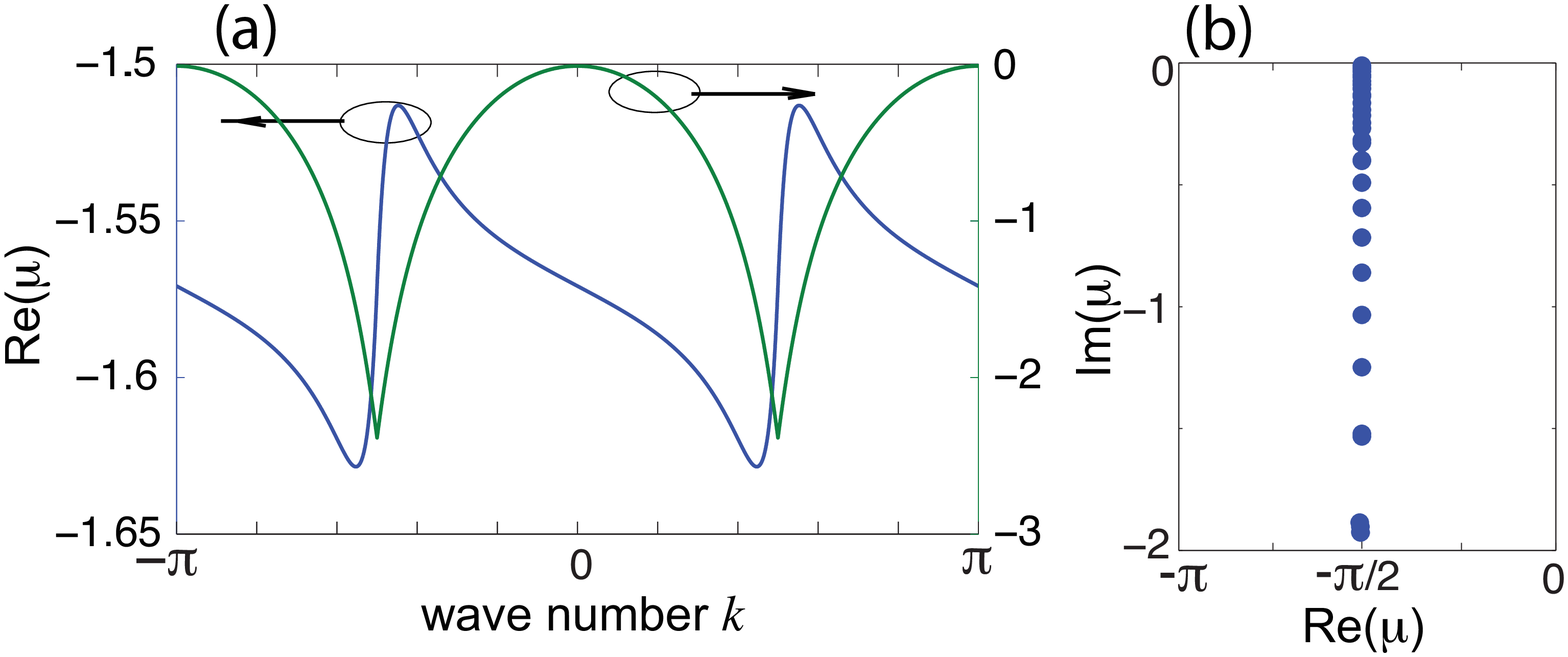}} \caption{ 
(Color online) (a) Lyapunov exponent $\mu$ versus Bloch wave number $k$ under PBC for a stochastic light walk with the random variable $\xi^{/m)}$ uniformly distributed in the range $(-\Delta,\Delta)$. Parameter values are $J=\Delta=0.5$. (b) Lyapunov exponents in a lattice comprising 40 sites under OBC.}
\end{figure} 
     \begin{figure}[htbp]
\centerline{\includegraphics[width=8.7cm]{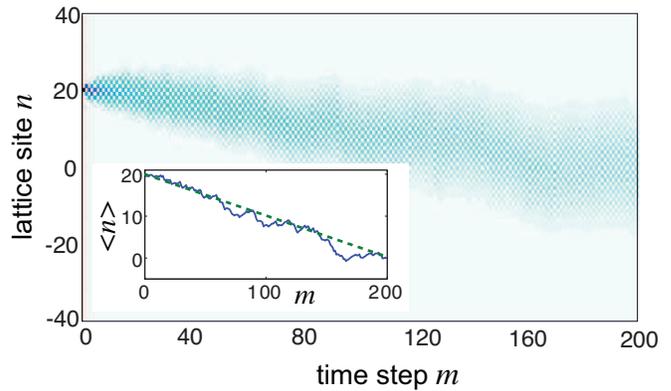}} \caption{ 
(Color online) Non-Hermitian drift induced by stochastic fluctuations with uniform distribution $f(\xi)$ for parameter values $J=\Delta=0.5$. 
The figure shows, on a pseudo color map, the behavior of the normalized light intensity $|a_n^{(m)}|^2 / \sum_n |a_n^{(m)}|^2$ versus time step $m$ for a given realization of the stochastic process.
The lattice is initially excited in the site $n=20$. The inset depicts the behavior of center of mass $\langle n \rangle$ of light excitation in the lattice versus $m$. The dashed curve in the inset corresponds to a constant drift at the velocity $v_g$ predicted by the Lyapunov exponent  analysis [Eq.(12)].}
\end{figure} 
  \begin{eqnarray}
  \mu(k) & = & -\frac{\pi}{2}+k +i( \log J-1)   \\
  & +&  \frac{i}{2\Delta} [1+\Delta + \exp(2ik)] \log [1+ \Delta + \exp(2ik)] \nonumber \\
  &- & \frac{i}{2\Delta} [1-\Delta + \exp(2ik)] \log [1- \Delta+ \exp(2ik)].  \nonumber
  \end{eqnarray}
A typical behavior of the real and imaginary parts of the Lyapunov exponent $\mu(k)$ versus $k$ is shown in Fig.4(a). The stochastic skin effect is clearly revealed after computation of the Lyapunov exponents in a finite lattice with OBC [Fig.4(b)]. Like in the time-independent case, the Lyapunov exponent under PBC is point-gapped and describes a closed loop in complex plane, indicating the topological origin of the stochastic skin effect as measured by the non-vanishing winding number $w=(1/2 \pi i) \int d \log{( \det (\mu(k)-E_B)})$ for any base energy $E_B$ inside the loop \cite{r3,r35}. However, under OBC the Lyapunov exponents collapse along a segment.
Under PBC, the largest growth rate is attained ay $k=0, \pi$, with a corresponding non-vanishing and negative value of the drift velocity $v_g= (d {\rm Re}(\mu) /dk)$, given by
\begin{equation}
v_g=1- \frac{\log(2+ \Delta)-\log(2-\Delta)}{\Delta}.
\end{equation}
Note that $v_g \rightarrow 0$ as $\Delta \rightarrow 0$, i.e. as the fluctuations vanish, indicating that the drift is driven by the stochastic term. The asymptotic behavior of 
$a_n^{(m)}$ at large values of $m$ is dominated by the spectral components with wave numbers near $k=0, \pi$ showing the largest growth rate, corresponding to a non-vanishing drift velocity. Hence, fluctuation-induced unidirectional light flow is expected, even thought on average the left/right hopping amplitudes remain equal. An example of stochastic-induced unidirectional light flow, as obtained for a given realization of the stochastic process
$\xi^{(m)}$, is shown in Fig.5. The direction of the drift can be reversed by just considering the fluctuation in $J_L$ rather than in $J_R$, i.e. by assuming [compare with Eq.(6)]
 \begin{equation}
 J_R=J \; , \; \; J_L= J(1+\xi^{(m)}).
 \end{equation}
We mention that above results hold for other distributions $f(\xi)$ of the stochastic process $\xi^{(m)}$, for example for a normal distribution or even for the simplest case of the Bernoulli distribution, where $\xi^{(m)}$ can assume only two values.\\
\\
       \begin{figure*}[htbp] 
\centerline{\includegraphics[width=17.5cm]{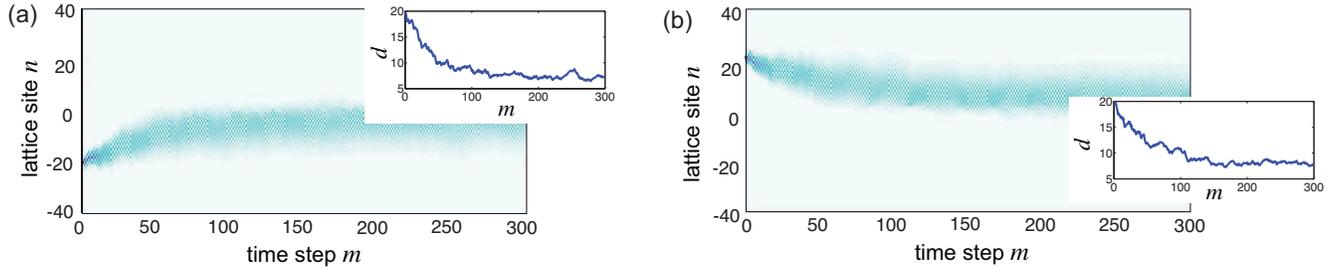}} \caption{ 
(Color online) Stochastic funneling of light at an interface for a uniform distribution $f(\xi)$ of the stochastic process $\xi^{(m)}$ and for parameter values $J=\Delta=0.5$. 
The two panels shows, on a pseudo color map, the behavior of the normalized light intensity $|a_n^{(m)}|^2 / \sum_n |a_n^{(m)}|^2$ versus time step $m$ for a given realization of the stochastic process
corresponding to single-site excitation of the site $n=-20$ in (a), and $n=20$ in (b). The inset depicts the behavior of distance $d$ of light excitation from the interface $n=0$ versus time step $m$. The distance $d$ is defined by the relation
$d(m)= \left\{ \sum_n n^2 |a_n^{(m)}|^2 / ( \sum_n |a_n^{(m)}|^2) \right\}^{1/2}$.}
\end{figure*} 
In the presence of a lattice boundary or an interface where the direction of the drift is reversed in the two half-lines, we expect accumulation of light excitation at the boundary or at the interface, an effect that can be referred to as {\em stochastic} non-Hermitian skin effect.  
However, it should be noted that, since we are dealing with a time-dependent system with stochastic dynamics, the concept of eigenvectors cannot be introduced and thus in our case the skin effect refers to accumulation of excitation at the edges or interfaces in the dynamics, rather than  condensation of eigenvectors. To illustrate the concept of stochastic non-Hermitian skin effect, let us consider an interface separating two half-lines with different fluctuations, namely
  \begin{eqnarray}
 J_L=J \; , \; \; J_R= J(1+\xi^{(m)}) \; \;\; n \geq 0 \\
 J_R=J \; , \; \; J_L= J(1+\xi^{(m)}) \;\;\; n<0
 \end{eqnarray}
 so that the drift velocity $v_g$ changes direction in the two half-lines and pushes any initial light excitation toward the interface. This effect realizes the stochastic version of light funneling shown in Fig.3 and recently demonstrated in \cite{r12}. Figure 6 shows an example of stochastic light funneling for two different initial conditions, clearly indicating the condensation of light excitation at the interface induced by the fluctuations.\\
\\
{\it Conclusions}
In this work we predicted a kind of non-Hermitian skin effect which is driven solely by stochastic temporal fluctuations in the system. Such a counterintuitive effect has been illustrated by discussing funneling of light pulses in fiber mesh lattices. While on average left and right hopping in the lattice remains symmetric, fluctuations can still induce unidirectional flow of excitation and its accumulation at system edges or interfaces. Like for the ordinary skin effect, the stochastic skin effect is rooted in the point-gap topology of the Lyapunov exponents under PBC. As we limited to consider skin effect in lattices induced by temporal fluctuations that maintain discrete translational spatial invariance, a similar effect is expected to arise when considering spatial fluctuations in the system, which destroy translational invariance \cite{r25}. Our results shed new light on stochastic non-Hermitian transport and non-Hermitian skin effect, and are expected to motivate further investigations in different areas of physics.\\
 \\
 The author acknowledges the
Spanish State Research Agency, through the Severo Ochoa
and Maria de Maeztu Program for Centers and Units of
Excellence in R\&D (Grant No. MDM-2017-0711).\\
 \\
The author declares no conflicts of interest.\\
\\


\begin{thebibliography}{99}

\bibitem{r1}
N. Hatano and D. R. Nelson, Phys. Rev. Lett. {\bf 77},
570 (1996).
\bibitem{r2}
 N. Hatano and D. R. Nelson, Phys. Rev. B {\bf 58}, 8384 (1998).
\bibitem{r3}
Z. Gong, Y. Ashida, K. Kawabata, K. Takasan, S.
Higashikawa, and M. Ueda, Phys. Rev. X {\bf 8}, 031079 (2018).
\bibitem{r4}
S. Longhi, D. Gatti, and G. Della Valle, Phys. Rev B {\bf 92}, 094204 (2015).
\bibitem{r5}
S. Longhi, D. Gatti, and G. Della Valle, Sci. Rep. {\bf 5},
13376 (2015).
\bibitem{r6}
X. Z. Zhang and Z. Song, Ann. Phys. {\bf 339}, 109 (2013).
\bibitem{r6b}
S Longhi, 
Opt. Lett. {\bf 40}, 1278 (2015).
\bibitem{r7}
A. Metelmann and A. A. Clerk, Phys.
Rev. X {\bf 5}, 021025 (2015).
\bibitem{r8}
S. Longhi, Ann. Phys. {\bf 530}, 1800023 (2018).
\bibitem{r10}
S Longhi and L Feng, 
APL Photonics {\bf 3}, 060802 (2018).
\bibitem{r11}
C. Li, L. Jin and Z. Song, 
Phys. Rev. A {\bf 95}, 022125 (2017).
\bibitem{r12}
S. Weidemann, M. Kremer, T. Helbig, T. Hofmann, A. Stegmaier, M. Greiter, R. Thomale, and A. Szameit, Science {\bf 368}, 311 (2020).
\bibitem{r13}
Z. Zhang, X. Qiao, B. Midya, K. Liu, J. Sun, T. Wu, W. Liu, R. Agarwal, J.M. Jornet, S. Longhi, N.M. Litchinitser, and L. Feng, Science {\bf 368}, 760 (2020).
\bibitem{r14}
S. Yao and Z. Wang, Phys. Rev. Lett. {\bf 121}, 086803 (2018).
\bibitem{r15}
 C.H. Lee and R. Thomale, Phys. Rev. B {\bf 99}, 201103 (2019).
\bibitem{r16}
F.K. Kunst, E. Edvardsson, J.C. Budich, and E.J. Bergholtz, Phys. Rev. Lett. {\bf 121}, 026808 (2018).
\bibitem{r17}
 V. M. Martinez Alvarez, J. E. Barrios Vargas, and L. E. F. Foa Torres, Phys. Rev. B {\bf 97}, 121401 (2018).
\bibitem{r18}
T. E. Lee, Phys. Rev. Lett. {\bf 116}, 133903 (2016).
\bibitem{r19}
 L. Xiao, T. Deng, K. Wang, G. Zhu, Z. Wang, W. Yi, and P. Xue, Nature Phys. {\bf 16}, 761 (2020).
\bibitem{r20}
T. Helbig, T. Hofmann, S. Imhof, M. Abdelghany, T. Kiessling, L.W. Molenkamp, C. H. Lee, A. Szameit, M. Greiter, and R. Thomale, Nature Phys. {\bf 16}, 747 (2020).
\bibitem{r21}
A. Ghatak, M. Brandenbourger, J. van Wezel, and C. Coulais, arXiv:1907.11619 (2019).
\bibitem{r22}
S. Longhi,
Opt. Lett. {\bf 44}, 5804 (2019).
\bibitem{r23}
S. Longhi, 
Phys. Rev. Res. {\bf 1}, 023013 (2019).
\bibitem{r23b}
K. Wang, X. Qiu, L. Xiao, X. Zhan, Z. Bian, W, Yi, and P. Xue,
Phys. Rev. Lett. {\bf 122}, 020501 (2019).
\bibitem{r24}
S. Longhi, 
Phys. Rev. Lett. {\bf 122}, 237601 (2019).
\bibitem{r25}
J. Claes and T.L. Hughes, arXiv:2007.03738v1 (2020).
\bibitem{r25b}
C.R. Doering, W. Horsthemke, and J. Riordan, 
Phys. Rev. Lett. {\bf 72}, 2984 (1994).
\bibitem{r26}
S. Derevyanko, Sci. Rep. {\bf 9}, 12883 (2019).
\bibitem{r27}
A. Schreiber, K. N. Cassemiro, V. Potocek, A. Gabris, P. J. Mosley, E. Andersson, I. Jex, and Ch. Silberhorn, Phys. Rev. Lett. {\bf 104}, 050502 (2010).
\bibitem{r28}
A. Schreiber, K. N. Cassemiro, V. Potocek, A. Gabris, I. Jex, and Ch. Silberhorn, Phys. Rev. Lett. {\bf 106}, 180403 (2011).
\bibitem{r29}
A. Regensburger, C. Bersch, B. Hinrichs, G. Onishchukov, A. Schreiber, C. Silberhorn, and U. Peschel, Phys. Rev. Lett. {\bf 107}, 233902 (2011).
\bibitem{r30}
A. Regensburger, C. Bersch, M.-A. Miri, G. Onishchukov, D.N. Christodoulides, and U. Peschel, Nature {\bf 488}, 167-171 (2012).
\bibitem{r31}
M. Wimmer, H.M. Price, I. Carusotto, and U. Peschel, Nature Phys. {\bf 13},  545 (2017).
\bibitem{r32}
I. D. Vatnik, A. Tikan, G. Onishchukov, D.V. Churkin, and A.A. Sukhorukov,  Sci. Rep. {\bf 7}, 4301 (2017).
\bibitem{r35}
 N. Okuma, K. Kawabata, K. Shiozaki, and M. Sato, Phys. Rev. Lett. {\bf 124}, 086801 (2020).














 
\end{thebibliography}
\end{document}